\documentclass[twocolumn,nofootinbib,amsmath,amssymb]{revtex4}
\usepackage{graphicx}
\usepackage{dcolumn}
\usepackage{bm}
\begin{document}
\title{Theoretical Study of Spin Observables in $\bm{pd}$ Elastic Scattering \\ at Energies $\bm{T_p = 800}$--$\bm{1000}$~MeV}
\author{M.~N.~Platonova}
\email{platonova@nucl-th.sinp.msu.ru}
\author{V.~I.~Kukulin}
\email{kukulin@nucl-th.sinp.msu.ru}
\affiliation{Skobeltsyn
Institute of Nuclear Physics, Lomonosov Moscow State University,
Moscow 119991, Russia}
\date{\today}
\begin{abstract}
Various spin observables (analyzing powers and spin-correlation parameters) in $pd$ elastic scattering at $T_p =
800$--$1000$~MeV are analyzed within the framework of the refined
Glauber model. The theoretical model uses as input
spin-dependent $NN$ amplitudes obtained from the most recent
partial-wave analysis and also takes into account the deuteron $D$
wave and charge-exchange effects. Predictions of the refined
Glauber model are compared with the existing experimental data.
Reasonable agreement between the theoretical calculations and
experimental data at low momentum transfers $|t| \lesssim
0.2$~(GeV/$c)^2$ is found for all observables considered. Moderate discrepancies found in this region are shown to be likely due to uncertainties in the input $NN$ amplitudes.
Qualitative agreement at higher momentum transfers is also found
for most observables except the tensor ones with mixed $x$ and $z$ polarization components. Possible reasons for observed deviations of the model calculations from the data at $|t|
> 0.2$~(GeV/$c)^2$ are discussed.
\end{abstract}
%
%
\maketitle
\section{\label{sec:intro} Introduction}
Proton-deuteron elastic scattering being the simplest
nucleon-nucleus collision process is well suited for studying the
basic properties of nuclear force. Due to small number of active
particles, $pd$ elastic scattering at intermediate energies is
widely used for testing different theoretical models of $2N$ and
$3N$ interaction, including non-standard mechanisms such as
production of baryon and dibaryon resonances. While $pd$ elastic
scattering has been extensively studied experimentally and
theoretically during more than 50 years, the new data are still
being accumulated, thus extending the existing database and
calling for suitable theoretical approaches for their
interpretation.

A large portion of the new precise experimental data on
intermediate-energy $pd$ scattering (both elastic and inelastic)
have come from the Cooler Synchrocyclotron at the J{\"u}lich FZ
(COSY) (see the dedicated review~\cite{Wilkin17}). One should
mention here the very recent COSY-ANKE measurements of the
small-angle $dp$ elastic differential cross section at equivalent
proton energies $T_p$ around $900$~MeV~\cite{Fritzsch18}, the
deuteron analyzing powers at $T_p = 600$ and
$1135$~MeV~\cite{Mchedlishvili18}, and the proton analyzing power
in $pd$ elastic scattering at $T_p = 796$~MeV and five higher
energies from $1600$ to $2400$~MeV~\cite{Barsov18}. Among the
other recent data collected after 2010 are the RIKEN measurements
of the deuteron analyzing powers at $T_p = 250$ and
$294$~MeV~\cite{Sekiguchi11,Sekiguchi14} and Dubna data on the
small-angle differential cross section and deuteron vector
analyzing power at $T_p = 1000$~MeV~\cite{Glagolev12}, the
large-angle differential cross section at $T_p$ from $500$ to
$900$~MeV~\cite{Terekhin17,Terekhin17-1} and $T_p =
1250$~MeV~\cite{Kurilkin12} and deuteron analyzing powers at $T_p
= 440$~MeV~\cite{Kurilkin12-1}.

The $pd$ and $dp$ elastic scattering data at energies $T_p <
350$~MeV can be analyzed theoretically by exact solution of the
three-body Faddeev equations with a realistic $NN$ potential as
input~\cite{Gloeckle96}. The theoretical calculations strongly
deviate from experiment at large scattering angles at $T_p
> 100$~MeV, and this deviation is generally attributed to the
effect of $3N$ forces~\cite{Sekiguchi11,Sekiguchi14}. However
inclusion of conventional $3N$ forces based on production of
intermediate $\Delta$ isobar removes the observed discrepancies
only partially. This is a deep puzzle intimately related to our
poor understanding of the short-range mechanisms of $NN$ and $3N$
interaction.

At higher energies, the situation is even more unclear. On the one
hand, numerous experiments (see, e.g., the works of Dubna and
Saclay groups~\cite{Komarov72,Azhgirei97,Azhgirei98,Punjabi95,Arvieux83,Arvieux84,Berthet82})
have revealed interesting structures in large-angle (in
particular, backward) $dp$ elastic scattering which indicate
manifestation of the non-nucleon, i.e., isobar and possibly
dibaryon, degrees of freedom (see also~\cite{Uzikov98}). On the
other hand, the existing high-precision $NN$ potentials cannot be
used in the region of $T_p > 350$~MeV, and hence the solution of
Faddeev equations becomes unreliable (and also too complicated).
So, one does not possess an exact theoretical treatment of $pd$
scattering in the GeV energy region even for two-body interactions
and low momentum transfers. In this area, one generally applies
the approximate multiple-scattering schemes, which usually take
into account single and double scattering between the proton and
nucleons in the deuteron and use $NN$ amplitudes as input. The
most famous one is the Glauber diffraction
theory~\cite{Glauber55,Franco66} which is a high-energy and
small-momentum transfer approximation to the exact
multiple-scattering series. There are also several more
sophisticated approaches (e.g., the relativistic
multiple-scattering
theory~\cite{Alberi82,Bleszynski86,Bleszynski87} or the more
recent one~\cite{Ladygina08,Ladygina09}), which include various
corrections to the Glauber approximation (mainly due to off-shell
and relativistic effects), however still restricted to single and
double scattering. Hence, these theoretical schemes essentially
account for the two lowest iterations of the Faddeev equations,
with inclusion of relativistic effects and some additional
mechanisms important for large-angle scattering. While such
approaches are much more involved than the Glauber diffraction
theory, they should be applied carefully, bearing in mind that
various corrections to the Glauber $pd$ scattering amplitude tend
to cancel each other, e.g., the off-shell part of the
double-scattering term (omitted in the Glauber model)
substantially cancels the contributions of higher order
rescattering terms~\cite{Harrington69}.

Thus, before considering any corrections to the Glauber theory, it
is worth developing the most accurate theoretical model within the
framework of the Glauber approximation, taking into account the
recent progress in describing $NN$ scattering and deuteron
properties. That was the motivation of elaborating the refined
Glauber model in Refs.~\cite{PKPRC10,PKYAF10}, which is
essentially the conventional Glauber diffraction model extended to
incorporate spin degrees of freedom. The refined version uses as
input all ten helicity $pp$ and $np$ amplitudes obtained from the
modern partial-wave analysis (PWA) of the George Washington
University group (SAID)~\cite{Arndt07,SAID-URL} and employs the
deuteron $S$- and $D$-wave functions derived in the high-precision
$NN$-potential models (e.g., CD-Bonn~\cite{Machleidt01}). The
model also takes into account the process of double charge
exchange, which is a manifestation of isospin dependence of the
general $NN$ amplitude. It should be noted that while the Glauber
approach cannot help resolve the numerous discrepancies found in
the large-angle $pd$ scattering, it gives an accurate theoretical
treatment of the small-angle region at intermediate energies.
Thus, given the high precision of experimental data and reliable
$NN$ input, even the small deviations between the Glauber model
calculations and experiment can be considered as indication of
some non-trivial effects, related to $3N$ forces and non-nucleon
degrees of freedom.

The generalized scheme~\cite{PKPRC10,PKYAF10} allowed for the first time to analyze within the diffraction model not only unpolarized cross sections, but also various
spin-dependent observables, which are much more sensitive to the
tiny details of $pd$ interaction. In previous
works~\cite{PKPRC10,PKYAF10} we calculated the proton and deuteron
analyzing powers at energies $T_p = 250$ and $440$ MeV and some of the deuteron analyzing powers at $1000$~MeV
and obtained very good agreement with existing experimental data
at transferred momenta squared $|t| < 0.3$--$0.4$~(GeV/$c)^2$. For
two lower energies, we also found surprisingly good agreement with
the exact three-body calculations based on solution of the Faddeev
equations in the same momentum-transfer region~\cite{PKPRC10}. In
the recent work~\cite{Temerbayev15} our formalism was applied to
$pd$ elastic scattering at $T_p = 135$ and $200$~MeV. A
very good description of experimental data on differential cross
sections, analyzing powers and some of the spin-correlation
parameters was achieved at these energies, though at first glance $T_p = 135$ MeV seems to be very low for application of the (high-energy) Glauber approach. The recent data on the vector and tensor deuteron analyzing
powers in $dp$ elastic scattering at the equivalent proton
energies $T_p = 600$ and $1135$~MeV~\cite{Mchedlishvili18} were
also very well described by our model, with somewhat better
agreement at higher energy.
Besides that, the model was quite successfully applied to the
antiproton-deuteron scattering at energies ranging from $50$ to
$300$~MeV~\cite{Uzikov13}.

The very recent application of the refined Glauber model was the
analysis of the new high-precision COSY-ANKE
data~\cite{Fritzsch18} on the $dp$ elastic differential cross
section taken at $T_p$ between $882.2$ and $918.3$~MeV in the
transferred momentum range $0.08 < |t| < 0.26$~(GeV/$c)^2$. In
these calculations, we used as input the recently updated $NN$ PWA
of the SAID group~\cite{Workman16} which gives somewhat different
$NN$ amplitudes than the older solution~\cite{Arndt07} at $T_p >
500$~MeV. Very good agreement between the theoretical calculation
at $T_p = 900$~MeV and experimental data was found at low momentum
transfers $|t| < 0.2$~(GeV/$c)^2$~\cite{Fritzsch18}. At higher
$|t|$ values, the theoretical calculation was shown to
significantly underestimate the data. Quite surprisingly, the
range of applicability of the refined Glauber model turned out to
be smaller than that found previously for lower energies (i.e.,
$|t| < 0.3$--$0.4$~(GeV/$c)^2$ at $T_p = 250$ and
$440$~MeV)~\cite{PKPRC10}. To establish the origin of this discrepancy,
it is important to consider also the
spin observables which can readily be calculated in the
refined Glauber model at the same energies.

Unfortunately, the only new data on $pd$ elastic spin observables
in the energy range $T_p = 800$--$1000$~MeV collected after 2010
are the above-mentioned data on the proton analyzing
power~\cite{Barsov18} and on the deuteron vector analyzing
power~\cite{Glagolev12}, the latter data having rather large
uncertainties. On the other hand, there exist a rich set of older
experimental data at $T_p = 800$~MeV, including all proton and
deuteron analyzing powers and a number of spin-correlation
parameters and spin-transfer
coefficients~\cite{Winkelmann80,Irom83,Adams88,Rahbar87,HajiSaied87,Ghazikhanian91,Igo88,Guelmez92,Sun85}.
These numerous $800$-MeV data have not yet been analyzed within
the Glauber model.

Thus, the aim of this work is to study a large number of spin observables in
$pd$ elastic scattering at energies $T_p = 800$--$1000$~MeV within
the refined Glauber model and compare the results with existing
experimental data. While previous works~\cite{PKPRC10,PKYAF10,Mchedlishvili18} dealt with some of the deuteron analyzing powers only (in the GeV energy region), here all proton and deuteron analyzing powers and also spin-correlation parameters are considered. This study will give a comprehensive test of
the model and provide a theoretical basis for the future experiments on $pd$ elastic spin observables
at these and higher energies. Such experiments with polarized proton and deuteron beams are planned at, e.g., the NICA-SPD facility under construction at JINR (Dubna, Russia)~\cite{SPD-NICA}.

The paper is organized as follows. In Sec.~\ref{sec:model} we briefly
outline the basic theoretical formalism. In Sec.~\ref{sec:results} we
present the results of the calculations and compare
them with experimental data. In Sec.~\ref{sec:discuss} we discuss the origin of discrepancies found between the calculations and the data. We conclude in Sec.~\ref{sec:summary}.
Some details on transformation of the amplitudes and observables between different notations are given in Appendix.

\section{\label{sec:model} Theoretical model}
The full formalism of the refined Glauber model was derived in
Refs.~\cite{PKPRC10,PKYAF10}. Here we briefly remind the basic
formulas for amplitudes and give expressions for observables which
were not considered previously.

The $pd$ elastic scattering amplitude $M$ in the Glauber model is
the sum of the single- and double-scattering terms which are
expressed through $pp$ and $pn$ amplitudes $M_p$ and $M_n$ and the
deuteron wave function $\Psi_d$:
\begin{eqnarray}
\label{msd} M({\bf q}) \!\!&=&\!\! M^{(s)}({\bf q}) + M^{(d)}({\bf
q}), \\
\label{ms}
 M^{(s)}({\bf q}) \!\!&=&\!\! \int \! d^{3}r e^{i{\bf q}\cdot{\bf r}/2}
\Psi_d({\bf r}) \bigl[M_n({\bf q}) + M_p({\bf q})\bigr]
 \Psi_d({\bf r}), \\
\label{md} M^{(d)}({\bf q}) \!\!&=&\!\! \frac{i}{4 \pi^{3/2}} \!
\int \! d^{2}q' \!\! \int \! d^{3}r e^{i{\bf q'}\cdot{\bf r}}
\Psi_d({\bf r}) \bigl[M_n({\bf q_2}) M_p({\bf
q_1}) \nonumber \\
\!\!& &\!\! + M_p({\bf q_2}) M_n({\bf q_1}) \!-\! M_c({\bf q_2})
M_c({\bf q_1}) \bigr] \Psi_d({\bf r}),
\end{eqnarray}
${\bf q}$ being the overall 3-momentum transfer (so that $t =
-q^2$ in the center-of-mass system), ${\bf q_1} = {\bf q}/2 - {\bf
q'}$ and ${\bf q_2} = {\bf q}/2 + {\bf q'}$ --- the momenta
transferred in collisions with individual target nucleons, and
$M_c({\bf q}) = M_n({\bf q}) - M_p({\bf q})$ --- the amplitude of
the charge-exchange process $pn \to np$. \\

The amplitude $M$ is expanded into $12$ terms invariant under
space and time reflections which are constructed from the scalar
products of the unit vectors $\hat{k} = ({\bf p}+{\bf p'})/|{\bf
p}+{\bf p'}|, \ \ \hat{q} = ({\bf p}-{\bf p'})/|{\bf p}-{\bf p'}|,
\ \ \hat{n} = \hat{k} \times \hat{q}$ forming the right-handed
system (${\bf p}$ and ${\bf p'}$ being the momenta of the incident
and outgoing proton, respectively) and spin vectors of the proton
($\frac{1}{2}{\bm \sigma}$) and deuteron ({\bf S}). The
coefficients of the expansion are the invariant amplitudes
$A_i(q)$, $i = 1,\ldots,12$, viz.
\begin{eqnarray}
\label{ma}
   M[{\bf p},{\bf  q}; {\bm \sigma}, {\bf  S}] \!&=&\! \bigl(A_1 + A_2 \,{\bm
\sigma}\!\cdot\!\hat{n}\bigr) + \bigl(A_3 + A_4 \,{\bm
\sigma}\!\cdot\!\hat{n}\bigr)({\bf S}\!\cdot\!\hat{q})^2 \nonumber \\
& & + \bigl(A_5 + A_6\,{\bm \sigma}\!\cdot\!\hat{n}\bigr)({\bf
S}\!\cdot\!\hat{n})^2
+ A_7 \, {\bm \sigma}\!\cdot\!\hat{k}\,{\bf S}\!\cdot\!\hat{k} \nonumber \\
& & + A_{8} \,{\bm \sigma}\!\cdot\!\hat{q}\,({\bf
S}\!\cdot\!\hat{q}\,{\bf S}\!\cdot\!\hat{n}
+ {\bf S}\!\cdot\!\hat{n}\,{\bf S}\!\cdot\!\hat{q}) \nonumber \\
& & + \bigl(A_9 + A_{10}\,{\bm \sigma}\!\cdot\!\hat{n}\bigr){\bf
S}\!\cdot\!\hat{n}
+ A_{11}\,{\bm \sigma}\!\cdot\!\hat{q}\,{\bf S}\!\cdot\!\hat{q} \nonumber \\
& & + A_{12}\,{\bm \sigma}\!\cdot\!\hat{k}\,({\bf
S}\!\cdot\!\hat{k}\,{\bf S}\!\cdot\!\hat{n} + {\bf
S}\!\cdot\!\hat{n}\,{\bf S}\!\cdot\!\hat{k}).
\end{eqnarray}

Analogously, the $pp$ and $pn$ amplitudes $M_N$ ($N = p,n$) are
expanded into six terms:
\begin{eqnarray}
\label{mn}
 M_N[{\bf p}, {\bf q};{\bm\sigma},{\bm\sigma}_N] \!&=&\!
A_N + C_N\,{\bm\sigma}\!\cdot\!\hat{n} +
C'_N\,{\bm\sigma}_N\!\cdot\!\hat{n} \nonumber \\
 & & +
B_N\,{\bm\sigma}\!\cdot\!\hat{k}\,{\bm\sigma}_N\!\cdot\!\hat{k} \nonumber \\
 & & + \bigl(G_N +
 H_N\bigr)\,{\bm\sigma}\!\cdot\!\hat{q}\,{\bm\sigma}_N\!\cdot\!\hat{q} \nonumber \\
 & & + \bigl(G_N - H_N\bigr)\,{\bm\sigma}\!\cdot\!\hat{n}\,{\bm\sigma}_N\!\cdot\!\hat{n},
\end{eqnarray}
where ${\bm\sigma}$ and ${\bm\sigma}_N$ are the Pauli matrices
corresponding to the incident and target nucleons. For the
double-scattering term, the unit vectors $\hat{k},\hat{q},\hat{n}$
are defined separately for each individual $NN$ collision.

Following the initial approach of Franco and
Glauber~\cite{Franco66}, we define the $pd$ as well as $NN$
amplitudes in the laboratory frame, where scattering off nucleon
and deuteron can be easily related to each other. Thus, we
distinguish the amplitudes $C_N$ and $C'_N$ which, for small
scattering angles, are interrelated as $C'_N \approx C_N +
i(q/2m_N)A_N$~\cite{Sorensen79}.\footnote{Note the inverted sign
of the amplitudes $C_N$ and $C'_N$ compared to that in
Ref.~\cite{Sorensen79}, due to a different definition of the unit
vector $\hat{n}$. Note also that the imaginary unit was missed in
the last term of Eq.~(14) in Ref.~\cite{PKPRC10}.} One should
however bear in mind that the unit vectors $\hat{k}$ and $\hat{q}$
which are orthogonal in the center-of-mass frame, are
approximately orthogonal in the laboratory frame at small
scattering angles, where $p \approx p'$. Assuming $p = p'$ and,
consequently, $-t = q^2$ in the laboratory frame is consistent
with the fixed-scatterer approximation inherent to the Glauber
theory.\footnote{One should note that the above relations are
valid for arbitrary $q$ not only in the center-of-mass, but also
in the Breit frame which is convenient to use for describing $pd$
(as well as $ed$) scattering when going beyond the fixed-scatterer
approximation. At small transferred momenta, the Breit frame
almost coincides with the laboratory frame.} The neglect of the
deuteron recoil energy is justified when it is small compared to
the momentum transfer, i.e., when $q^2/4m_d^2 \ll 1$.

The final formulae for all $pd$ amplitudes $A_1$--$A_{12}$
expressed in terms of the $NN$ amplitudes $A_N$, $B_N$, $C_N$,
$C'_N$, $G_N$ and $H_N$ ($N = n,p$) and the deuteron monopole and
quadrupole form factors are to be found in
Refs.~\cite{PKPRC10,PKYAF10}. In the present calculations, we used
the $NN$ amplitudes corresponding to
the recent SAID PWA solution SM16~\cite{Workman16,SAID-URL} and the high-precision CD-Bonn deuteron wave function~\cite{Machleidt01}.
Both the $NN$ amplitudes and the deuteron wave function
were parameterized as sums of five Gaussian
terms as described in Refs.~\cite{PKPRC10,PKYAF10}.

We complete this section with the definitions for $pd$ elastic
observables. The differential cross section is related to the
amplitude $M$ as
\begin{equation}
\label{dsm}
 d\sigma/dt =
\frac{1}{6}\,{\rm Sp}\,\left(MM^+\right).
\end{equation}
The general polarization observable is defined as (see,
e.g.,~\cite{Ghazikhanian91})
\begin{equation}
\label{gpo} C(\alpha,\beta,\gamma,\delta) = \frac{{\rm
Sp}\,\left(M\sigma_{\alpha}S_{\beta}M^+\sigma_{\gamma}S_{\delta}\right)}{{\rm
Sp}\,\left(MM^+\right)},
\end{equation}
where $\alpha = \{x,y,z\}$ and $\gamma = \{x',y',z'\}$ correspond
to the polarization of the initial and final proton, while $\beta
= \{x,y,z,xx,yy,zz,xy,zy,xz\}$ and $\delta =
\{x',y',z',x'x',y'y',z'z',x'y',z'y',x'z'\}$
--- to the (vector or tensor) polarization of the initial and final deuteron.
For the tensor values of $\beta \equiv \beta_1\beta_2$
($\beta_1,\beta_2 = \{x,y,z\}$), $S_{\beta}$ means the quadrupole
operator $S_{\beta_1\beta_2} = \frac{3}{2}(S_{\beta_1}S_{\beta_2}
+ S_{\beta_2}S_{\beta_1}) - 2\delta_{\beta_1\beta_2}$ (the same
holds for the index $\delta$). Each of the four indices
$\alpha,\beta,\gamma,\delta$ can also be zero, which means that
the respective particle has no definite polarization.

In this paper we deal with the observables corresponding to the
polarized initial particles, i.e., beam and target. These are the
proton (vector) analyzing powers $A_{\alpha}^p \equiv
C(\alpha,0,0,0)$, the deuteron vector and tensor analyzing powers
$A_{\alpha}^d \equiv C(0,\alpha,0,0)$ and $A_{\beta} \equiv
C(0,\beta,0,0)$ for $\beta = \beta_1\beta_2$, and the vector and
tensor spin-correlation parameters ${C_{\alpha,\beta}} \equiv
C(\alpha,\beta,0,0)$, all with non-zero values of $\alpha$ and
$\beta$.\footnote{All observables of the type
$C(\alpha,\beta,0,0)$ are sometimes denoted in literature as
$A_{\alpha\beta}$ (see, e.g.,~\cite{Adams88}); in this case,
spin-correlation parameters are called correlated analyzing
powers, though this notation is far less common.}

If to define the coordinate system $xyz =
\{\hat{q}\hat{n}\hat{k}\}$, the $pd$ elastic observables can be
readily expressed in terms of the invariant amplitudes
$A_1$--$A_{12}$ (see Eq.~(\ref{ma})). The explicit formulas for
the differential cross section and all vector and tensor analyzing
powers were presented in
Refs.~\cite{PKPRC10,PKYAF10}.\footnote{Note that the ``-'' sign of
$A_{xz}$ was accidentally dropped in Eq.~4 of
Refs.~\cite{PKPRC10} and~\cite{PKYAF10}.}
Here we add the formulas for non-vanishing vector and tensor
spin-correlation parameters:
\begin{eqnarray}
\label{scp}
  C_{y,y} \!&=&\! 2\,{\rm Re}\bigl[(2A_1^* + A_3^*
+ 2A_5^*)A_{10} \nonumber \\
\!& &\! + (2A_2^* + A_4^* + 2A_6^*)A_9 - A_7^*A_{11} - A_8^*A_{12}\bigr]/\Sigma, \nonumber \\
  C_{x,x} \!&=&\! 2\,{\rm Re}\bigl[(2A_1^* + 2A_3^*
+ A_5^*)A_{11} - A_6^*A_{12} \nonumber \\
\!& &\! - A_7^*A_{10} + A_8^*A_9\bigr]/\Sigma, \nonumber \\
  C_{z,x} \!&=&\! 2\,{\rm Im}\bigl[(2A_2^* + 2A_4^*
+ A_6^*)A_{11} - A_5^*A_{12} \nonumber \\
\!& &\! + A_7^*A_9 - A_8^*A_{10}\bigr]/\Sigma, \nonumber \\
  C_{x,z} \!&=&\! -2\,{\rm Im}\bigl[(2A_2^* + A_4^*
+ A_6^*)A_7 + (A_3^* - A_5^*)A_8 \nonumber \\
\!& &\! - A_9^*A_{11} + A_{10}^*A_{12}\bigr]/\Sigma, \nonumber \\
  C_{z,z} \!&=&\! 2\,{\rm Re}\bigl[(2A_1^* + A_3^*
+ A_5^*)A_7 + (A_4^* - A_6^*)A_8 \nonumber \\
\!& &\! - A_{10}^*A_{11} + A_9^*A_{12}\bigr]/\Sigma, \nonumber \\
  C_{y,yy} \!&=&\! 2\,{\rm Re}\bigl[A_1^*(2A_6 - A_4) + A_3^*(A_6 - A_2 - A_4) \nonumber \\
\!& &\! + A_5^*(2A_2 + A_4 + 2A_6) - 3A_7^*A_8 + 2A_9^*A_{10} \nonumber \\
\!& &\! - 3A_{11}^*A_{12}\bigr]/\Sigma, \nonumber \\
  C_{y,xx} \!&=&\! 2\,{\rm Re}\bigl[A_1^*(2A_4 - A_6) + A_3^*(2A_2 + 2A_4 + A_6) \nonumber \\
\!& &\! + A_5^*(A_4 - A_2 - A_6) + 3A_7^*A_8 - A_9^*A_{10} \bigr]/\Sigma, \nonumber \\
  C_{y,xz} \!&=&\! -3\,{\rm Im}\bigl[A_3^*A_{10} + A_4^*A_9 + A_7^*A_{11} + A_8^*A_{12}\bigr]/\Sigma, \nonumber \\
  C_{y,zz} \!&=&\! -C_{y,yy} - C_{y,xx}, \nonumber \\
  C_{x,xy} \!&=&\! 3\,{\rm Re}\bigl[(2A_1^* + A_3^*
+ A_5^*)A_8 + (A_4^* - A_6^*)A_7 \nonumber \\
\!& &\! + A_9^*A_{11} - A_{10}^*A_{12}\bigr]/\Sigma, \nonumber \\
  C_{z,xy} \!&=&\! 3\,{\rm Im}\bigl[(2A_2^* + A_4^*
+ A_6^*)A_8 + (A_3^* - A_5^*)A_7 \nonumber \\
\!& &\! - A_9^*A_{12} + A_{10}^*A_{11}\bigr]/\Sigma, \nonumber \\
  C_{x,zy} \!&=&\! -3\,{\rm Im}\bigl[(2A_2^* + 2A_4^*
+ A_6^*)A_{12} - A_5^*A_{11} \nonumber \\
\!& &\! + A_8^*A_9 - A_7^*A_{10}\bigr]/\Sigma, \nonumber \\
  C_{z,zy} \!&=&\! 3\,{\rm Re}\bigl[(2A_1^* + 2A_3^*
+ A_5^*)A_{12} - A_6^*A_{11} \nonumber \\
\!& &\! + A_7^*A_9 - A_8^*A_{10}\bigr]/\Sigma,
\end{eqnarray}
where
\begin{eqnarray}
\label{si} \Sigma \!&=&\! 3\,d\sigma/dt = 3\bigl(|A_1|^2 +
|A_2|^2\bigr) + 2\Bigl(\sum\limits_{i=3}^{12}{|A_i|^2} \nonumber \\
& & + {\rm Re}\bigl[2A_1^*(A_3 + A_5) + 2A_2^*(A_4 + A_6) \nonumber \\
& & + A_3^*A_5 + A_4^*A_6\bigr]\Bigr).
\end{eqnarray}

To compare our model calculations with experimental data,
we have to transform all $pd$ observables to the Madison frame
which is conventionally used in experiments, i.e., $xyz =
\{\hat{S}\hat{N}\hat{L}\}$, where $\hat{L} = \hat{p}$, $\hat{N} =
\widehat{p \times p'}$, and $\hat{S}=\hat{N}\times\hat{L}$.\footnote{A
lot of measurements were actually performed on $dp$ rather than
$pd$ scattering. The Madison frame for $dp$ scattering is related
to that for $pd$ scattering by reflection of two axes $x \to -x$,
$z \to -z$, which does not affect the definition of observables.}
In fact, the $pd$ invariant amplitudes can be defined directly in
the Madison frame as was done in Ref.~\cite{Temerbayev15}.
However, we prefer to use the advantages of the
$\{\hat{q}\hat{n}\hat{k}\}$ coordinate system conventional for the
Glauber model calculations. The symmetry between the initial and
final states allows to easily apply $T$-invariance which leads to
the 12 independent amplitudes, rather than 18 dependent ones which
have to be dealt with in the Madison frame. The observables
obtained in terms of these 12 amplitudes can then be readily
transformed to any other coordinate system.

The Madison frame is related to the $\{\hat{q}\hat{n}\hat{k}\}$
system by the rotation in the scattering plane ($xz$) by the half
scattering angle $\theta/2$, and the reflection of the normal
($y$) axis, viz.
\begin{equation}
\hat{S} = -a \hat{q} + b \hat{k}, \quad \hat{N} = -\hat{n}, \quad
\hat{L} = b \hat{q} + a \hat{k},
\end{equation}
where $a = {\rm cos}(\theta/2)$, $b = {\rm sin}(\theta/2)$. So,
the transformed observables involving just the $y$-axis will only
change their sign ($A_y^p$, $A_y^d$ and $C_{y,yy}$) or even remain
unchanged ($A_{yy}$ and $C_{y,y}$). Other observables containing
$x$ and/or $z$ indices would slightly change their behavior at
small scattering angles. The explicit transformation rules for the
observables considered here are given in Appendix.

\section{\label{sec:results} Results}
\begin{figure}
\resizebox{1.0\columnwidth}{!}{\includegraphics{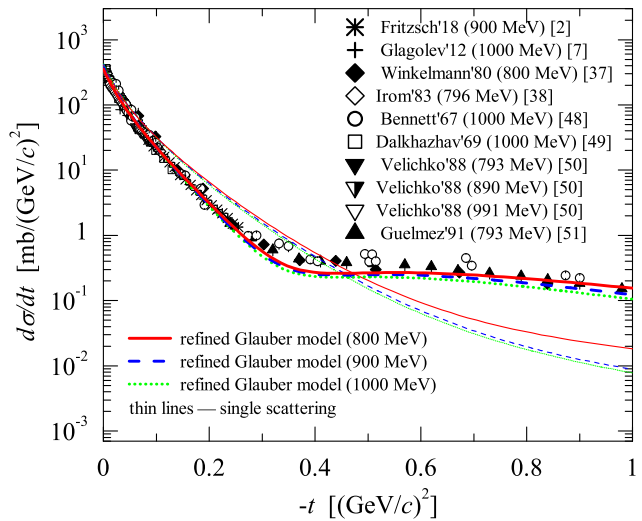}}
\caption{ (Color online) Differential cross sections of $pd$
($dp$) elastic scattering at the incident (equivalent) proton
energies $T_p = 800$--$1000$~MeV. Solid (red), dashed (blue) and dotted (green) lines show the refined
Glauber model calculations at $T_p = 800$, $900$ and
$1000$~MeV, respectively, with the input $NN$ amplitudes corresponding to the SAID PWA solution SM16~\cite{Workman16,SAID-URL}. Thin lines of the same type (color) show
the single-scattering contribution only. Theoretical calculations
are compared with the existing experimental database at $T_p =
793$--$1000$~MeV~\cite{Glagolev12,Fritzsch18,Winkelmann80,
Irom83,Bennett67,Dalkhazhav69,Velichko88,Guelmez91}.} \label{fig:dsdt}
\end{figure}

\begin{figure}
\begin{center}
\resizebox{1.0\columnwidth}{!}{\includegraphics{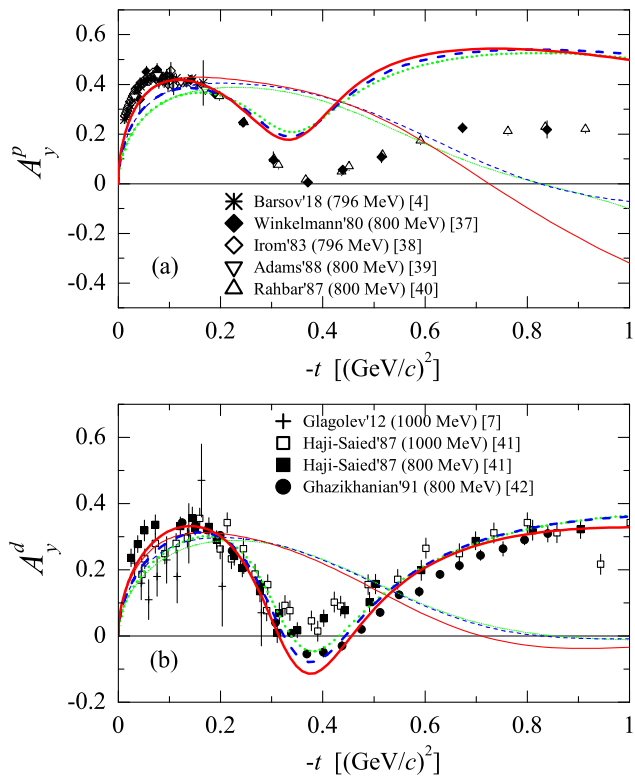}}
\end{center}
\caption{ (Color online) Proton (a) and deuteron (b) vector
analyzing powers in $pd$ ($dp$) elastic scattering at the incident
(equivalent) proton energies $T_p = 800$--$1000$~MeV. The meaning
of lines is the same as in Fig.~\ref{fig:dsdt}. Experimental data are
taken from
Refs.~\cite{Winkelmann80,Irom83,Adams88,Rahbar87,HajiSaied87,Ghazikhanian91,Glagolev12,Barsov18}.}
\label{fig:vap}
\end{figure}

\begin{figure}
\begin{center}
\resizebox{1.0\columnwidth}{!}{\includegraphics{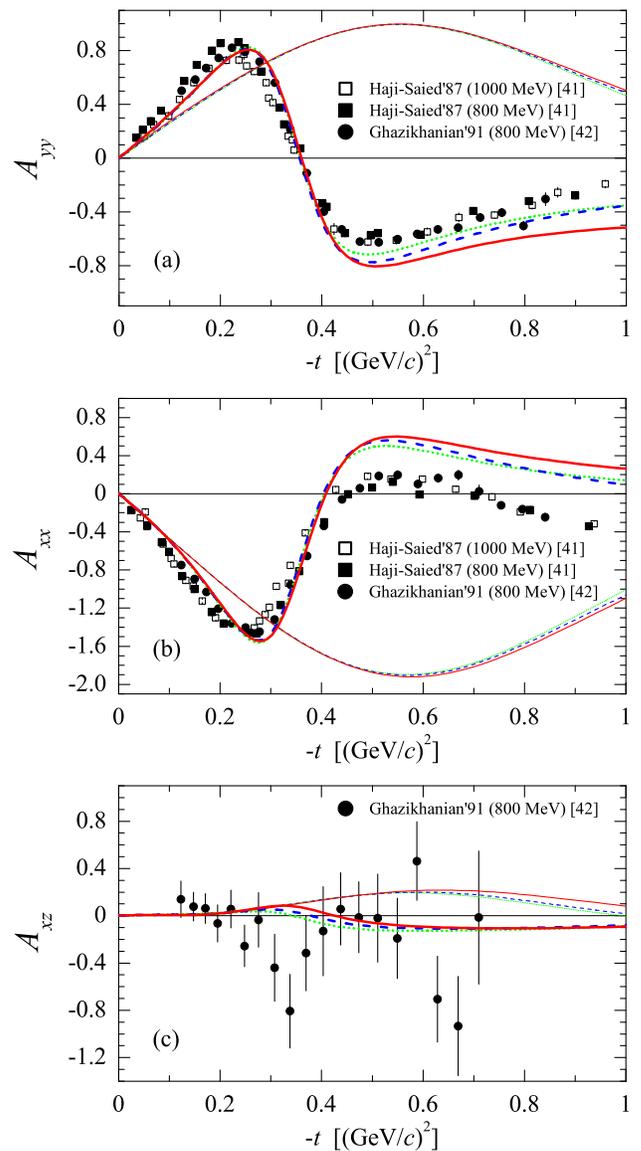}}
\end{center}
\caption{ (Color online) Deuteron tensor analyzing powers in $dp$
elastic scattering at the equivalent proton energies $T_p =
800$--$1000$~MeV. The meaning of lines is the same as in
Fig.~\ref{fig:dsdt}. Experimental data are taken from
Refs.~\cite{HajiSaied87,Ghazikhanian91}.} \label{fig:tap}
\end{figure}

\begin{figure}
\begin{center}
\resizebox{1.0\columnwidth}{!}{\includegraphics{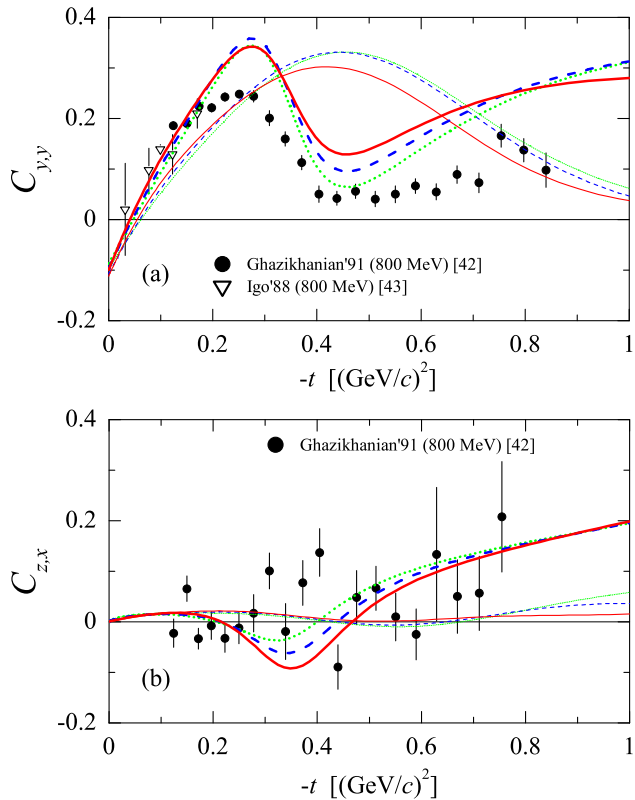}}
\end{center}
\caption{ (Color online) Vector spin-correlation parameters in
$pd$ ($dp$) elastic scattering at the incident (equivalent) proton
energies $T_p = 800$--$1000$~MeV. The meaning of lines is the same
as in Fig.~\ref{fig:dsdt}. Experimental data are taken from
Refs.~\cite{Ghazikhanian91,Igo88}.} \label{fig:vscp}
\end{figure}

\begin{figure}
\begin{center}
\resizebox{1.0\columnwidth}{!}{\includegraphics{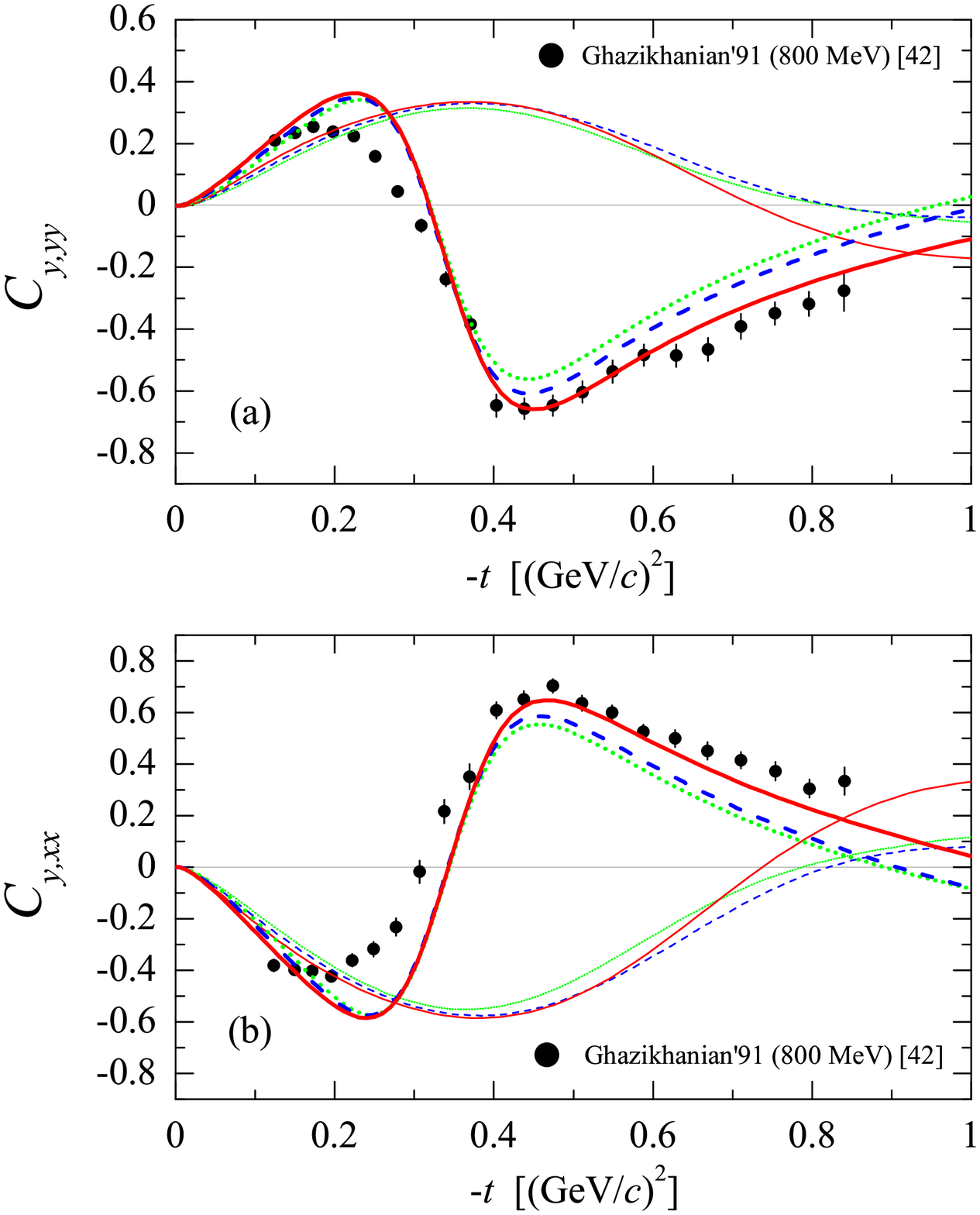}}
\resizebox{1.0\columnwidth}{!}{\includegraphics{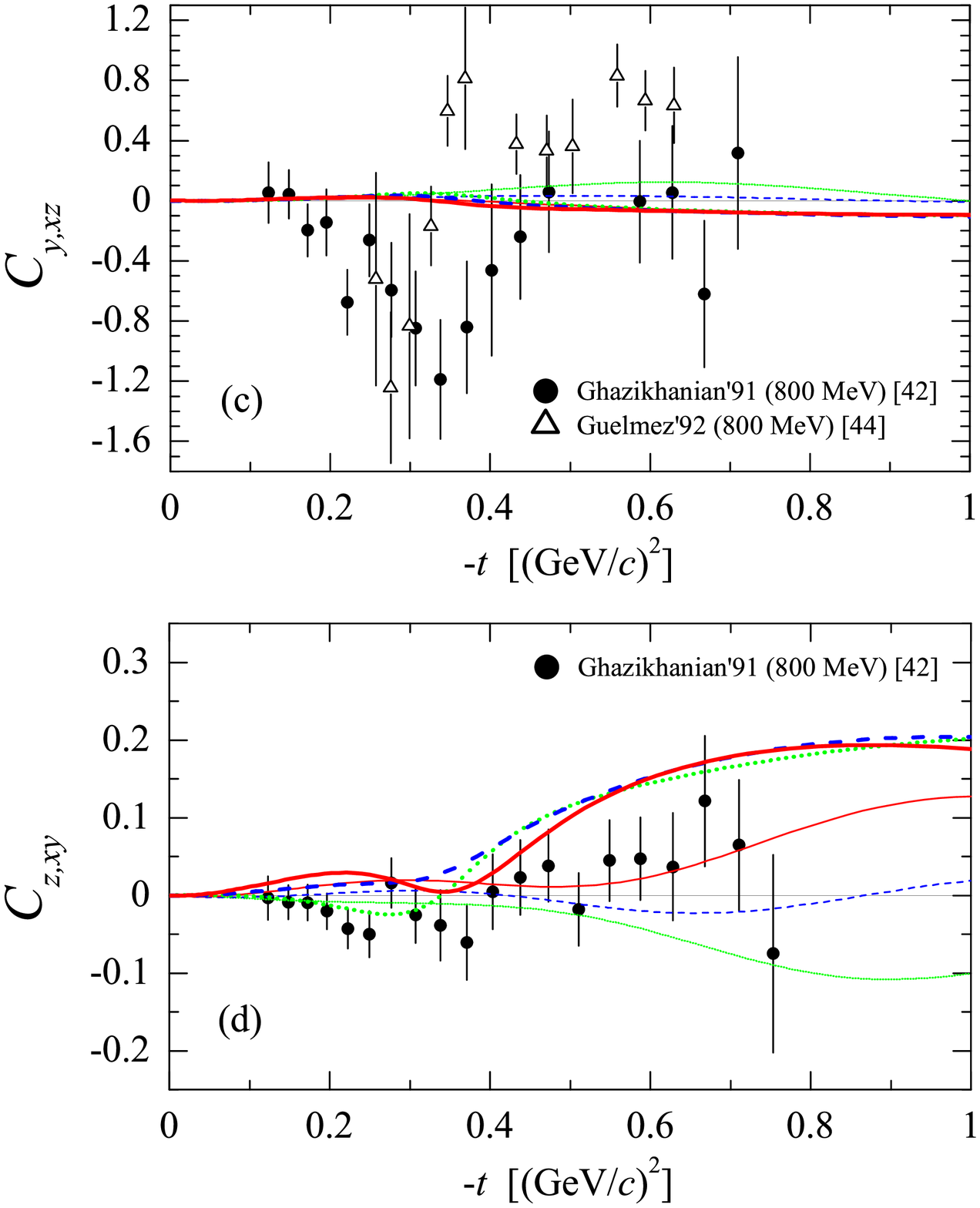}}
\end{center}
\caption{ (Color online) Tensor spin-correlation parameters in
$pd$ ($dp$) elastic scattering at the incident (equivalent) proton
energies $T_p = 800$--$1000$~MeV. The meaning of lines is the same
as in Fig.~\ref{fig:dsdt}. Experimental data are taken from
Refs.~\cite{Ghazikhanian91,Guelmez92}.} \label{fig:tscp}
\end{figure}

The results of the refined Glauber model calculations for $pd$
elastic differential cross section and various spin observables at
the proton energies $T_p = 800$, $900$ and $1000$~MeV are
presented in Figs.~\ref{fig:dsdt}--\ref{fig:tscp}. Our theoretical
predictions are compared with the experimental data available in
this energy region.\footnote{The available data include also $dp$
elastic scattering measurements at the incident deuteron energies
which are twice the proton energies considered here.} There are
numerous data on the unpolarized differential cross
section~\cite{Glagolev12,Fritzsch18,Winkelmann80,
Irom83,Bennett67,Dalkhazhav69,Velichko88,Guelmez91}, the proton
analyzing
power~\cite{Barsov18,Winkelmann80,Irom83,Adams88,Rahbar87},
deuteron analyzing
powers~\cite{Glagolev12,HajiSaied87,Ghazikhanian91},
spin-correlation
parameters~\cite{Adams88,Ghazikhanian91,Igo88,Guelmez92} and
spin-transfer
coefficients~\cite{Adams88,Rahbar87,Ghazikhanian91,Igo88,Guelmez92,Sun85}.
Most of these data were taken at $T_p = 800$~MeV. We plot here our
results for the differential cross section and 11 spin observables
(out of 22 needed for a complete experiment) measured in a broad
range of momentum transfers (including the low-momentum transfer
region), i.e., the proton and deuteron vector analyzing powers,
three deuteron tensor analyzing powers and six (out of 12)
spin-correlation parameters
--- two vector and four tensor ones --- measured in~\cite{Ghazikhanian91}.
All polarization observables are plotted in the Madison frame (see
Sec.~\ref{sec:model} and Appendix).

Our results for the differential cross section in comparison with
the existing experimental data at $T_p = 800$--$1000$~MeV are
presented in Fig.~\ref{fig:dsdt}. The contribution of single
scattering is also shown by thin lines.\footnote{The
single-scattering contribution is plotted in
Figs.~\ref{fig:dsdt}--\ref{fig:tscp} up to $|t| = 1$~(GeV/$c)^2$, though
the Gaussian parametrization~\cite{PKPRC10} used in calculations
reproduces $NN$ helicity amplitudes only for $q \leq
0.7$~GeV/$c$ ($|t| \leq 0.5$~(GeV/$c)^2$) and may deviate from them
at higher momentum transfers. Nevertheless, $pd$ elastic scattering at $|t|
> 0.35$~(GeV/$c)^2$ is dominated by the
double-scattering term, which contains $NN$ amplitudes in the
vicinity of $|t|/4$, so, the results of the full calculations correspond to the
correct $NN$ input until $|t| = 1$~(GeV/$c)^2$ and higher.} As is seen
from the figure (see also Ref.~\cite{Fritzsch18}), the refined
Glauber model calculations accurately describe the experimental
database at low momentum transfers~\footnote{Here and further on
we do not consider the region $0 \le |t| \lesssim 0.04$~(GeV/$c)^2$ where
Coulomb effects neglected in our model calculations are
significant in $pd$ scattering.} $|t| \leq 0.2$~(GeV/$c)^2$ and
then begin to deviate from the data, though remaining in the
qualitative agreement with them in the forward hemisphere.
It is also clearly seen from Fig.~\ref{fig:dsdt} that the energy dependence (the
increasing slope) of the calculated cross section as a function of
$|t|$, which arises mainly from the similar energy dependence of
the input $NN$ cross sections, is almost negligible at $|t| \leq
0.2$~(GeV/$c)^2$ but becomes clearly visible at higher $|t|$. So,
in the region where the Glauber model describes the data, the
cross section is almost energy-independent at $T_p =
800$--$1000$~MeV, and this is confirmed by the existing data.
However, a more accurate theoretical model should be used to study
the energy dependence of the $pd$ cross section at higher momentum
transfers.

The results for the vector and tensor analyzing powers are shown
in Figs.~\ref{fig:vap} and \ref{fig:tap}. Here we see again reasonable
(though not perfect) agreement with the data at $|t| \leq
0.2$~(GeV/$c)^2$. Some deviations from the data found in this
region are likely to be related to the uncertainties in the
input $NN$ amplitudes (see the discussion in Sec.~\ref{ssec:lowt}). One should also bear in mind the uncertainties present in the $pd$ data. Thus, if to compare two datasets~\cite{HajiSaied87} and
\cite{Ghazikhanian91} for the deuteron analyzing powers at $T_p =
800$~MeV (see Figs.~\ref{fig:vap} and \ref{fig:tap}), it is seen that the
latter data are described a bit better by our model calculations.
In fact, these data were obtained by the same group
as~\cite{HajiSaied87} and thus appear to be more precise, though
still having some uncertainty in the beam polarization. Further,
it is quite surprising that the proton
analyzing power $A_y^p$ at $|t| \leq
0.3$~(GeV/$c)^2$ is reproduced better at low energies ($440$,
$250$ and even $135$~MeV~\cite{PKPRC10,Temerbayev15}) than in the GeV region, though the applicability of the Glauber theory should improve with energy. The situation here is similar but even more drastic as for the differential cross section, and the reason probably lies in the uncertainties of the input $NN$ amplitudes rising with energy. In this connection, it would be very instructive to study the behavior of $A_y^p$ at $T_p \ge 1$ GeV theoretically, especially in view of the recently
published data~\cite{Barsov18} at $T_p$ from $1600$ to $2400$~MeV.
Unfortunately, these data cannot be presently analyzed by the
refined Glauber model, due to absence of the reliable $np$
PWA at energies $T_p >
1300$~MeV~\cite{Workman16}.

Figs.~\ref{fig:vap} and \ref{fig:tap} also show at least qualitative (or
even semiquantitative) agreement between the Glauber theory and
experiment at $|t| > 0.2$~(GeV/$c)^2$ for all vector and tensor
analyzing powers, except for the tensor one $A_{xz}$, which is
poorly reproduced at $0.2 < |t| < 0.4$~(GeV/$c)^2$. While the experimental $A_{xz}$ has a pronounced dip in this region, the refined Glauber model calculations give the smooth curve very close to zero. In fact, this $|t|$ region includes the transition between the dominance of the single- and
double-scattering terms, which is very sensitive to the tiny
details of the $pd$ scattering process, and especially to the
deuteron $D$-wave contribution (or, more precisely, to the quadrupole
deuteron form factor containing the interference between the $S$-
and $D$-wave components). So, the tensor analyzing power $A_{xz}$
appears to be highly sensitive to an accurate description of the
terms associated with the deuteron quadrupole form factor (see the discussion in Sec.~\ref{ssec:hight}).

The situation with the vector and tensor spin-correlation
parameters is quite similar to that with the analyzing powers. As
is seen from Figs.~\ref{fig:vscp} and \ref{fig:tscp}, all considered
spin-correlation parameters are described quite well at low momentum
transfers $|t| \leq 0.2$~(GeV/$c)^2$, and most of them
are described at least qualitatively at higher momentum transfers up to $|t| = 1$~(GeV/$c)^2$,.
Only the tensor spin-correlation parameter $C_{y,xz}$ which behaves similarly to the tensor analyzing power $A_{xz}$ is poorly reproduced in the region $0.2 < |t| < 0.5$~(GeV/$c)^2$,
likely for the same reason as $A_{xz}$.
Unfortunately, the most problematic observables with the mixed $x$ and $z$ polarization components
are also those having the largest experimental uncertainties. So,
to get a more clear picture of their theoretical description, it
would be highly desirable to measure these observables with better
statistics.

We further turn to the analysis of energy dependence found for
$pd$ spin observables in the region $T_p = 800$--$1000$~MeV. As is
seen from Figs.~\ref{fig:vap}--\ref{fig:tscp}, our calculations reveal
some energy dependence of the vector analyzing powers at low
momentum transfers, where the difference between the results at
$T_p = 800$ and $1000$~MeV is up to 25\%. This theoretical
prediction is qualitatively confirmed by the experimental
data~\cite{HajiSaied87} for $A_y^d$ shown in Fig.~\ref{fig:vap}. Quite
similar but weaker energy dependence was found for the vector
spin-correlation parameters. On the other hand, the tensor
analyzing powers and tensor spin-correlation parameters, except
for $C_{z,xy}$, were found to be almost independent of energy at
$|t|$. Since the energy dependence of observables calculated within the refined Glauber model comes mainly from the similar dependence of the input $NN$ amplitudes, the energy
independence is quite expectable for the tensor observables which are mostly
sensitive to the deuteron wave function.
On the other hand, the spin-correlation parameter $C_{z,xy}$ which was
found to be strongly energy dependent at low momentum
transfers provides a crucial test for the input $NN$ amplitudes
and their treatment in the model for $pd$ elastic scattering.

\section{\label{sec:discuss} Discussion}

\subsection{\label{ssec:lowt} Sensitivity of the low-momentum-transfer $\bm{pd}$ observables to the input $\bm{NN}$ amplitudes}

Influence of the $NN$ input on the results obtained within the
refined Glauber model for $pd$ observables is worth to be
discussed in detail.

First, we studied the deviations between the different SAID PWA
solutions by comparing the results for $pd$ differential
cross section and spin observables at $T_p = 1$~GeV obtained
with the use of three $NN$ PWA solutions: two recent
solutions SM16 (unweighted) and WF16 (weighted)~\cite{Workman16} and an
older one SP07~\cite{Arndt07} used in our previous
works~\cite{PKPRC10,PKYAF10}. The difference between these three solutions is almost negligible at
$T_p \leq 500$~MeV, but becomes visible at $T_p \simeq 1$~GeV~\cite{Workman16}.
We found that the predictions based on different
PWA solutions are almost indistinguishable at low momentum
transfers $|t| \leq 0.2$~(GeV/$c)^2$ and begin to slightly diverge at
higher $|t|$. However this divergence is considerably smaller than the
discrepancies between the theoretical results and experimental
data at $|t| > 0.2$~(GeV/$c)^2$. So, while it is natural to choose the most recent (unweighted) solution SM16~\cite{Workman16} for the Glauber model calculations, any one out of the three above PWA solutions may be used.

Second, though the SAID PWA is considered to be reliable in the region $T_p \leq 1$~GeV, there are some
discrepancies between the energy-dependent solutions and $NN$
experimental data. For instance, we found an underestimation of
the recent high-precision $A_y^p$ data in $pp$ scattering at $T_p =
796$~MeV~\cite{Bagdasarian14} by all SAID PWA solutions
starting from SP07~\cite{SAID-URL}, which at small scattering
angles (corresponding to $|t| \simeq 0.05$~(GeV/$c)^2$) amounts to
5--7\%. Some underestimation takes place also for the recent $pn$ $A_y^p$ data~\cite{Barsov18}, though there are much less datapoints than in the $pp$ $A_y^p$ measurement~\cite{Bagdasarian14}.
This deviation between the SAID PWA solutions and $pN$ $A_y^p$ data apparently leads to some underestimation of the $pd$ $A_y^p$ and $A_y^d$ data at small $|t|$ in our model calculations. Indeed, in small-angle $pd$ scattering at intermediate energies $A_y^d$ is approximately proportional to $A_y^p$ (with a coefficient of $2/3$)~\cite{Uzikov19}, and $A_y^p$ is in turn approximately equal to the average $A_y^p$ in $pp$ and $np$ scattering~\cite{Wilkin-pc}.

To test the impact of the input $NN$ amplitudes on the small-angle behavior of $pd$ observables, we tried to find a modification of the SAID $NN$ amplitudes which could simultaneously improve the description of both $NN$ and $pd$ observables at small momentum transfers. Since the analyzing powers in $pN$ scattering $A_y^p = -2{\rm Re}[(A_N^*+G_N^*-H_N^*)C_N]/(d\sigma_N/dt)$ (with the sign given for the Madison frame) are mostly sensitive to the interference between the central $A_N$ and spin-orbit $C_N$ amplitudes, we tried to fit the existing $pN$ $A_y^p$ data at $|t| < 0.5$~(GeV/$c)^2$ by adjusting the spin-orbit amplitudes $C_p$ and $C_n$. Simultaneously, we fitted the existing data on $pN$ differential cross sections $d\sigma_N/dt$ and spin-correlation parameters $A_{yy}=2{\rm Re}[A_N^*(G_N-H_N)-B_N^*(G_N+H_N)+|C_N|^2]/(d\sigma_N/dt)$ to fix the moduli of the amplitudes $C_N$ and also the $pn$ charge-exchange cross section (which contains $|C_n-C_p|^2$) to fix their relative phase. We also included in the fit the data on $A_y^p$ in $pd$ scattering.

\begin{figure}
\begin{center}
\resizebox{1.0\columnwidth}{!}{\includegraphics{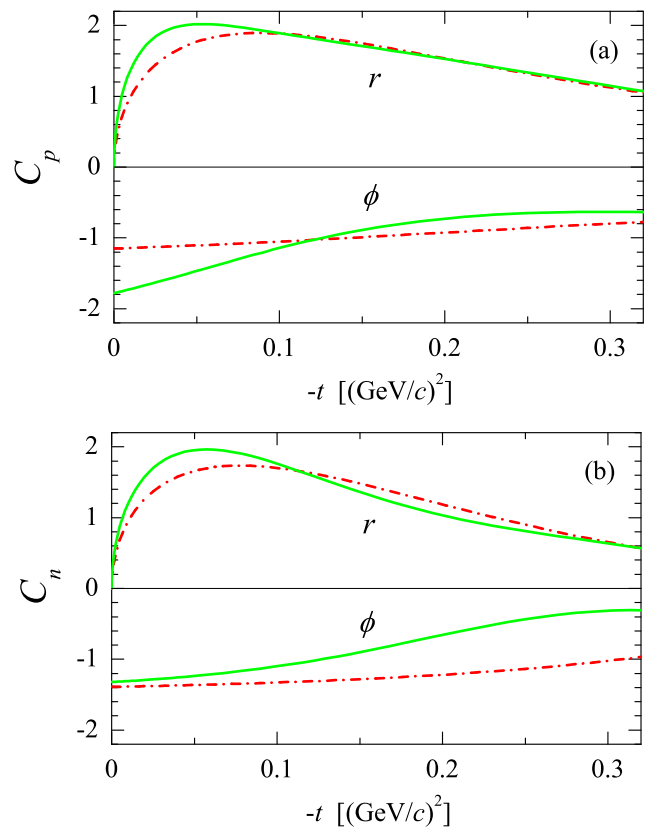}}
\end{center}
\caption{(Color online) Amplitudes $C_p$ ($a$) and $C_n$ ($b$) for $pp$ and $pn$ elastic scattering at $T_p=800$ MeV, respectively (see Eq.~(\ref{mn})), presented in the form $r e^{i\phi}$. Upper lines on each figure show the moduli $r$ [$\sqrt{\rm mb}/$GeV], and lower lines show the phases $\phi$ [rad.] of the amplitudes. Dash-dotted (red) lines correspond to the SAID PWA solution SM16~\cite{Workman16,SAID-URL}, while solid (green) lines show the modified $NN$ amplitudes which allow for a better description of $NN$ and $pd$ observables simultaneously (see text).} \label{fig:cpn}
\end{figure}

\begin{figure}
\begin{center}
\resizebox{1.0\columnwidth}{!}{\includegraphics{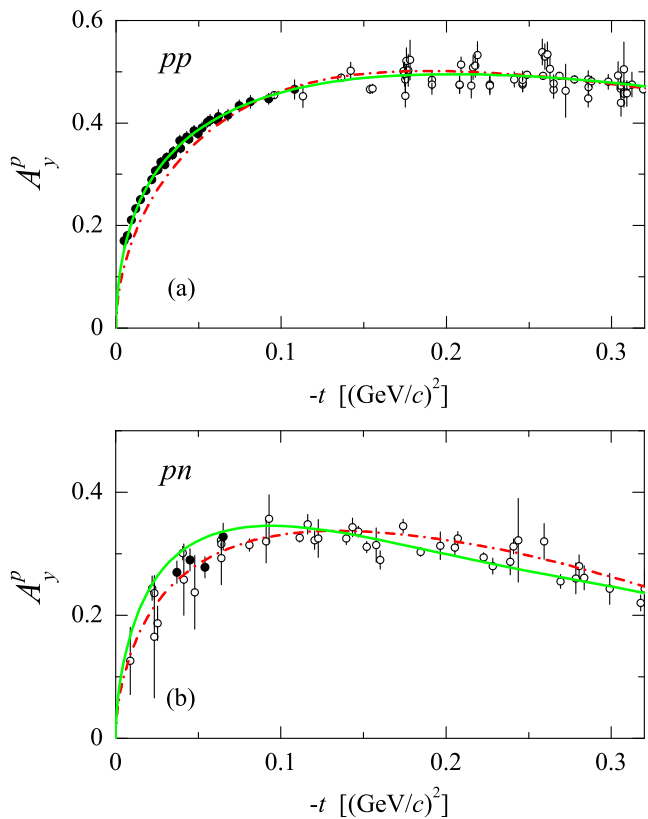}}
\end{center}
\caption{(Color online) Analyzing power $A_y^p$ in $pp$ (a) and $pn$ (b) elastic scattering at $T_p = 800$~MeV. Dash-dotted (red) lines correspond to the SAID PWA solution SM16~\cite{Workman16,SAID-URL}, while solid (green) lines show calculations with modified $NN$ amplitudes $C_p$ and $C_n$ (see Fig.~\ref{fig:cpn}). Filled circles show the recent ANKE-COSY experimental data at $T_p = 796$~MeV from Refs.~\cite{Bagdasarian14} ($pp$) and~\cite{Barsov18} ($pn$), and open circles show other (older) data at $T_p = 790$--$810$~MeV from the SAID database~\cite{SAID-URL}.}
\label{fig:ay-nn}
\end{figure}

\begin{figure}[!ht]
\begin{center}
\resizebox{0.95\columnwidth}{!}{\includegraphics{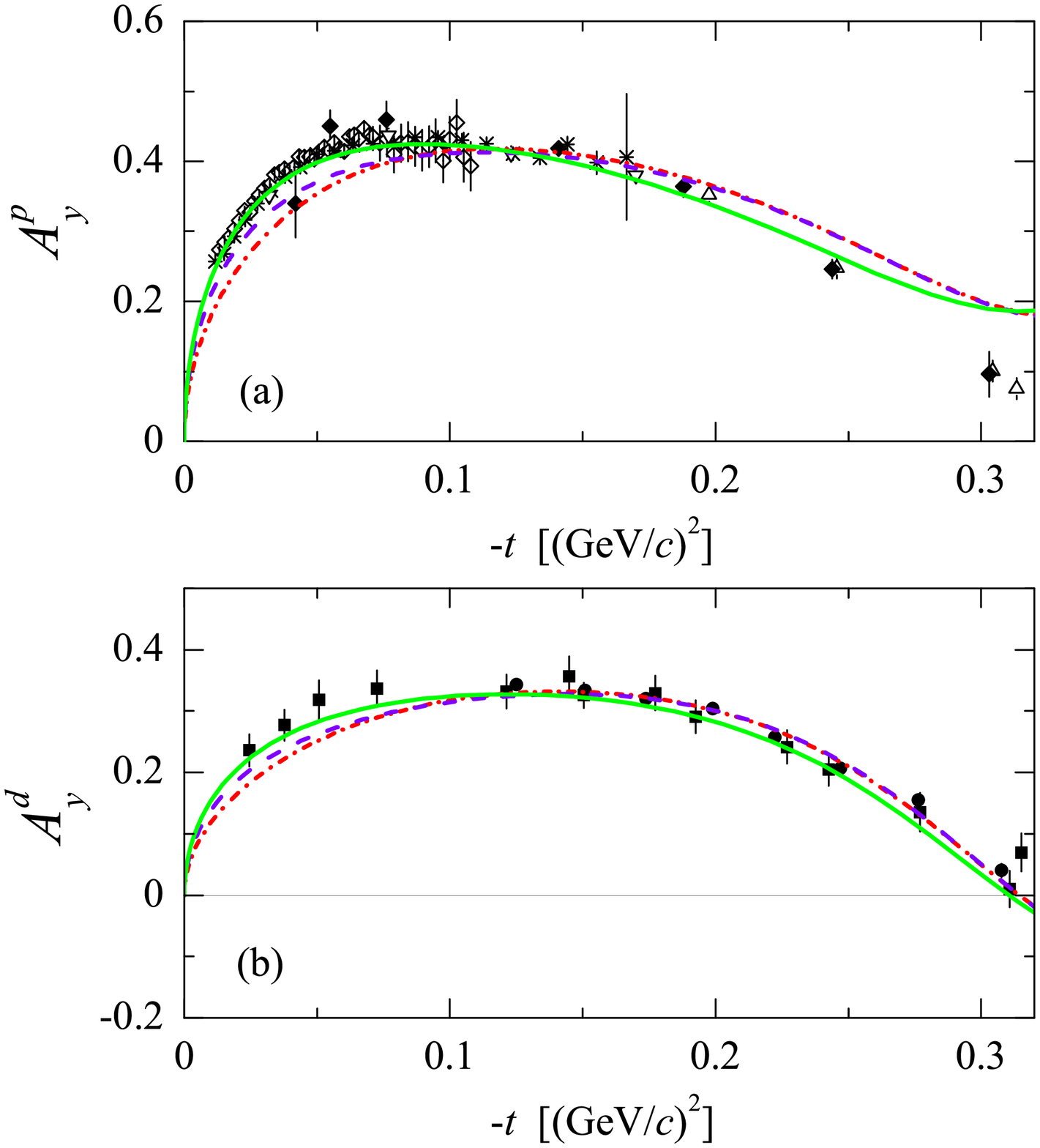}}
\resizebox{0.95\columnwidth}{!}{\includegraphics{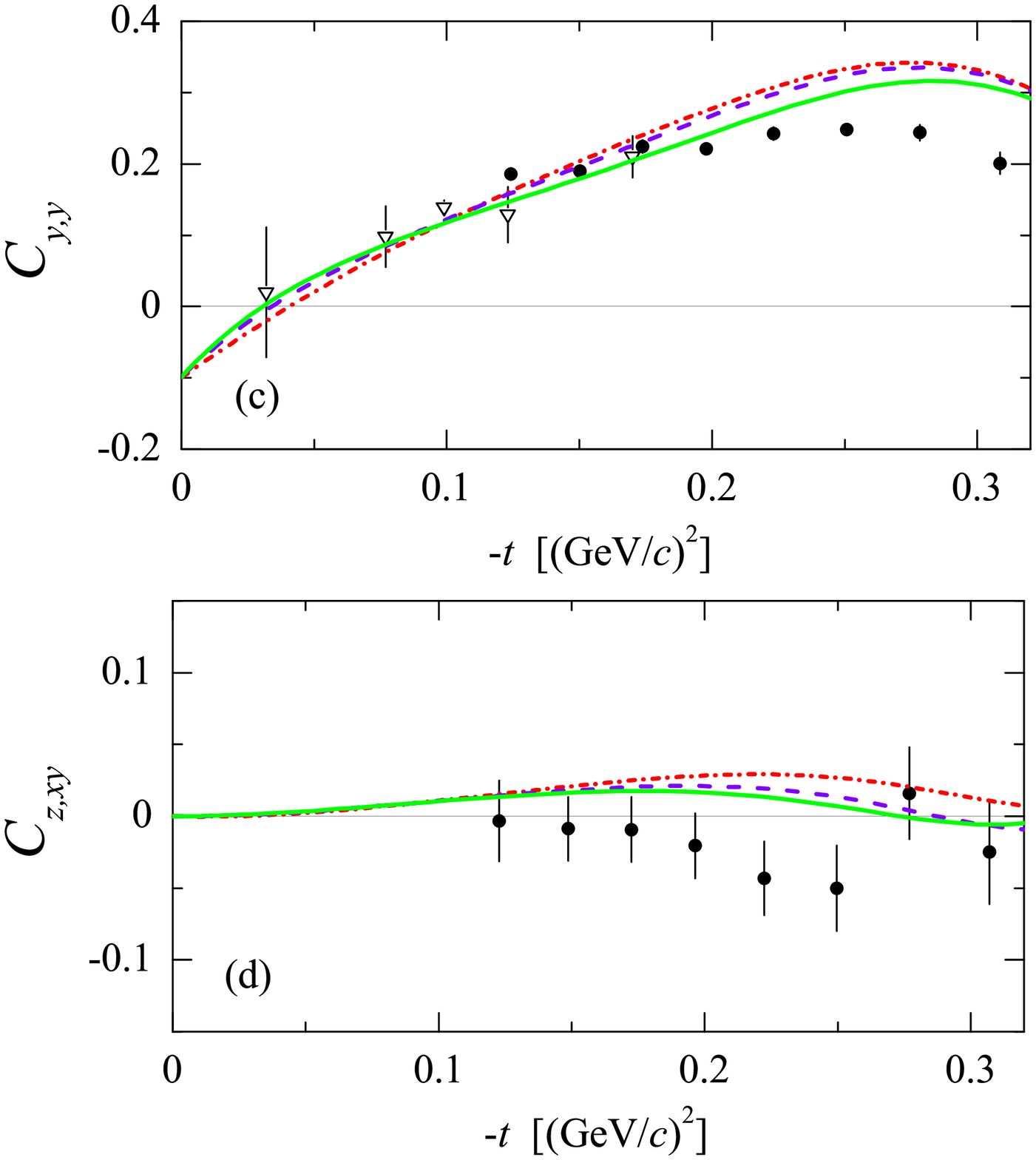}}
\end{center}
\caption{(Color online) Proton
analyzing power $A_y^p$ (a), deuteron
analyzing power $A_y^d$ (b) and vector spin-correlation parameter
$C_{y,y}$ (c) in $pd$ ($dp$) elastic scattering at the incident
(equivalent) proton energy $T_p = 800$~MeV. Dash-dotted (red), dashed (violet) and solid (green) lines show the refined Glauber model
calculations with $NN$ amplitudes corresponding to the SAID PWA solution SM16~\cite{Workman16,SAID-URL}, with the modified amplitude $C_p$ and with modified amplitudes $C_p$ and $C_n$ (see Fig.~\ref{fig:cpn}), respectively. Experimental data at $T_p = 796$ and $800$~MeV are the same as in Figs.~\ref{fig:vap}, \ref{fig:vscp} and~\ref{fig:tscp}.}
\label{fig:obs-mod}
\end{figure}

As a result, we found a modification of the spin-orbit $NN$ amplitudes (shown in Fig.~\ref{fig:cpn}) that improves significantly the description of small-angle $pp$ $A_y^p$ data (see Fig.~\ref{fig:ay-nn}) without worsening the description of other $NN$ data included in the fit, compared to the SAID SM16 solution. As is clearly seen from Fig.~\ref{fig:obs-mod}, the same modification allows for a significantly better description of both proton and deuteron analyzing powers in $pd$ elastic scattering within the refined Glauber model, whereas their ratio remains almost unchanged due to its weak sensitivity to the spin-orbit $NN$ amplitudes~\cite{Uzikov19}. Simultaneously, we achieve better agreement with the data for the vector spin-correlation parameter $C_{y,y}$ and, what is very important, for the tensor spin-correlation parameter $C_{z,xy}$, which is extremely sensitive to the input $NN$ amplitudes, in the region $|t| < 0.3$~(GeV/$c)^2$ (see Fig.~\ref{fig:obs-mod}). Other $pd$ observables considered here are reproduced at the same level of accuracy as with initial (SAID SM16) $NN$ amplitudes.

We also found some other discrepancies between the recent SAID PWA solutions and $NN$ data (e.g., a 10\% overestimation of the $pn$ forward and backward (charge-exchange) cross sections) which could not be removed by any modification of the spin-orbit $pn$ amplitude. Removing all these discrepancies requires a thorough revision of the SAID PWA which is not our task here. We have just shown the possibility to improve the description of small-angle $pd$ observables in the refined Glauber model by adjusting the input $NN$ amplitudes consistently with $NN$ experimental data. This result suggests that the remaining discrepancies for $pd$ observables at $|t| < 0.3$ (GeV/$c)^2$ could be removed by further refinement of the $NN$ input.

The ambiguity of the input $NN$ amplitudes
obtained from the PWA in the GeV energy region
is apparently related to experimental uncertainties in the $NN$ data which
affect the PWA solutions. While the uncertainties in the $pp$ or
$pn$ amplitudes can in principle be traced in the $NN$
experimental data, their estimation is much more nontrivial for
the sum of $pp$ and $pn$ amplitudes entering the single-scattering
term of the Glauber $pd$ amplitude, and even more complicated for
the bilinear combinations of $NN$ amplitudes entering the
double-scattering term. These problems with $NN$ amplitudes might as well be the
reason for the better description of $pd$
observables at $135$ and $250$ MeV than in the GeV region~\cite{PKPRC10,Temerbayev15}, since under (and
slightly above) the pion production threshold the exact unitarity
imposes more rigorous constraints on the $NN$ amplitudes obtained
by PWA.

At the same time, we were unable to find a modification of $NN$ amplitudes consistent with $NN$ data that could significantly reduce the discrepancies between our model calculations and $pd$ data at $|t| \ge 0.3$ (GeV/$c)^2$. So, the main reason for these discrepancies
is to be sought in the limitations of the Glauber model and
missing dynamical contributions.

\subsection{\label{ssec:hight} Possible reasons for discrepancies at higher momentum transfers}
In this subsection we discuss possible reasons for the
failure of the refined Glauber model in description of $pd$ elastic observables at $|t| > 0.2$--$0.3$ (GeV/$c)^2$. The gradually rising deviation between the data and theoretical calculations which are seen for the most $pd$ observables are quite expectable in view of the decreasing validity of the Glauber approach with momentum transfer. On the other hand, the severe discrepancies have been found for the tensor analyzing power $A_{xz}$ and spin-correlation parameter $C_{y,xz}$
in the single-to-double scattering
transition region ($0.2 < |t| < 0.5$~(GeV/$c)^2$), which is very sensitive to the deuteron $D$-wave contribution. These discrepancies appear to be not just the consequence of the low-momentum approximation. It should be
noted here that terms which connect the deuteron $D$-wave with the
product of two spin-dependent $NN$ amplitudes were neglected in
the double-scattering amplitude of our
model~\cite{PKPRC10,PKYAF10}. This approximation was justified by
the relative smallness of the spin-dependent $NN$ amplitudes in
comparison to the large spin-independent ones in the GeV energy region.
Though this assumption becomes less accurate as the
momentum transfer increases, it still works well in the
double-scattering amplitude where $NN$ amplitudes enter in the
vicinity of $|t|/4$. So, the small omitted terms can hardly give a
sizeable contribution to the observables in the considered momentum transfer region.

\begin{figure}
\begin{center}
\resizebox{0.9\columnwidth}{!}{\includegraphics{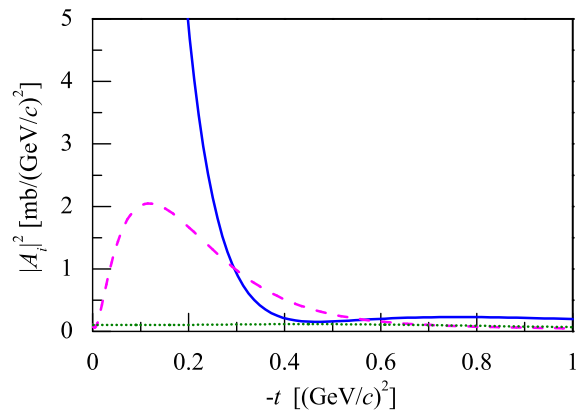}}
\end{center}
\caption{ (Color online) Moduli squared of the $pd$ elastic
scattering amplitudes $A_1$ (solid line), $A_3$ (dashed line) and
$A_5$ (dotted line) at $T_p = 800$~MeV calculated within the
refined Glauber model.} \label{fig:a135}
\end{figure}

The more detailed analysis shows that the dominant contribution to
$pd$ elastic scattering at GeV energies and small momentum transfers $|t| <
0.2$~(GeV/$c)^2$ comes from the largest invariant amplitude $A_1$.
At higher momentum transfers, the amplitudes $A_3$ (which
dominates at $0.3 < |t| < 0.55$~(GeV/$c)^2$) and $A_5$ also become
significant (see Fig.~\ref{fig:a135}). Other nine invariant amplitudes
are considerably smaller than $A_1$ at all momentum transfers. In
fact, all three above amplitudes include sizeable terms associated
with the deuteron quadrupole form factor, but these terms in $A_3$
and $A_5$ have an opposite sign as compared to that in $A_1$ (see
Table II in Ref.~\cite{PKPRC10}). Hence, they substantially cancel
each other in such observables as $A_y^d$ and $A_{yy}$ which
contain the combination $2A_1 + A_3 + 2A_5$, but not in $A_{xz}$
which contains a large contribution from $A_3$ not compensated by
$A_1$ (see Eq.~(4) in Ref.~\cite{PKPRC10}). So, $A_{xz}$ should be more sensitive to an accurate
description of the terms related to the deuteron quadrupole form
factor, than other analyzing powers. Quite similar conclusions could be drawn also for the tensor spin-correlation parameter $C_{y,xz}$.

It should be emphasized that the problem with description of the
tensor $pd$ observables appear to be tightly connected to a
long-standing problem with the deuteron quadrupole moment. In
fact, its experimental value is $Q^{\rm exp} = 0.2859(3)$~fm$^2$,
while the high-precision $NN$ potentials predict a bit lower
values $Q^{\rm theor} \simeq
0.270$--$0.280$~fm$^2$~\cite{Machleidt01}. This small (2--6\%)
discrepancy seems from first glance to be not very serious.
However it is likely related to the inaccuracy in short-range
behavior of the deuteron $D$-wave function where the error may be
large if to take into account the low probability of the $D$-wave
component in the deuteron. Thus this inaccuracy in the short-range
deuteron $D$ wave will translate to errors for the tensor
observables, such as $A_{xz}$ and $C_{z,xy}$. Hence, it would be very instructive
to study the sensitivity of these observables at $0.2 < |t| <
0.5$~(GeV/$c)^2$ to the behavior of the deuteron $D$-wave
short-range component. This however should be done beyond the
on-shell approximation for the double-scattering term, which
prevents the proper account of the short-range deuteron structure
in the Glauber theory.

On the other hand, the observed discrepancies might be related to some
dynamical mechanisms of the short-range nature, which were included neither in the Glauber model nor
in the more sophisticated multiple-scattering approaches. In fact,
even the fully spin-dependent relativistic multiple-scattering
theory~\cite{Bleszynski86} with parameters of the off-shell $NN$
amplitudes adjusted to fit the whole $pd$ elastic database at
$800$ MeV was unable to describe satisfactory the above tensor
observables~\cite{Ghazikhanian91}. So, one of the possible
candidates for the missing dynamical mechanisms is some kind of
$3N$ forces. While the conventional $3N$ force based on the
intermediate $\Delta$ excitation was shown to contribute
predominantly to the large-angle $pd$
scattering~\cite{Sekiguchi11,Sekiguchi14}, the non-conventional
three-body force which arise from the meson exchange between the
incident proton and the deuteron as a whole (i.e., the six-quark,
or dibaryon, component of the deuteron)~\cite{Kukulin04-1} might
give some sizeable corrections to the multiple-scattering
amplitude in the forward hemisphere~\cite{PKJPCS12}. Such a $3N$
force is tightly interrelated with the short-range $S$- and $D$-wave
components of the deuteron wave function, and it has also a strong
spin-orbit term, which can affect the description of polarization
observables sensitive to the spin-orbit interactions (such as
$A_y^p$). These interesting questions certainly deserve further
investigation and will be considered in our future work.

\section{Summary}
\label{sec:summary}

In this work we presented comparison between the predictions for
the differential cross section and various spin observables in
$pd$ elastic scattering in the GeV energy region given by the
refined Glauber model (with full account of spin degrees of
freedom) and experimental data. As an input in our model
calculations, fully realistic $NN$ helicity amplitudes obtained
from the most recent PWA (SAID) and the accurate deuteron $S$- and
$D$-wave functions derived within the high-precision $NN$ potential
model (CD-Bonn) were employed. While in our previous
works~\cite{PKPRC10,PKYAF10} only three deuteron analyzing powers at $T_p =
1$~GeV were considered, in this paper 11 polarization observables including all five analyzing powers and six spin-correlation parameters
at proton energies $T_p = 800$, $900$ and $1000$~MeV have been
calculated, thus providing a more complete picture.

As is clearly seen from Figs.~\ref{fig:dsdt}--\ref{fig:tscp}, our results
are in quite reasonable agreement with available experimental data
at transferred momenta squared $|t| \lesssim 0.2$ (GeV/$c)^2$ for
all observables considered. Some deviations from the data found in this region are likely to be due to uncertainties in the input $NN$ amplitudes arising from experimental uncertainties of $NN$ elastic scattering data and ambiguities in the PWA procedure above
the pion production threshold. In fact, we were able to remove some discrepancies between the SAID PWA solutions and $NN$ data at $T_p = 800$ MeV and simultaneously improve the description of a number of $pd$ spin observables at $|t| < 0.3$ (GeV/$c)^2$ by adjusting the spin-orbit $NN$ amplitudes. Though good agreement with the data at small momentum transfers is generally expectable for the Glauber theory, it has been achieved here for the first time for a large number of highly sensitive spin observables, thus providing a reliable theoretical basis for the future experiments with polarized proton and deuteron beams, which are planned, e.g., at the NICA-SPD facility at JINR, Dubna. Since the same formalism is straightforwardly applicable for the $\bar{p}d$ scattering, it can also serve as a tool for analysis of the future experiments with polarized antiproton beams, which are planned, e.g., within the FAIR project in GSI Darmstadt.

Though the description of the data by our model calculations clearly worsens with rising momentum transfer, we found at least qualitative agreement between theory and experiment for the most observables considered at $|t|$ up to 1~(GeV/$c)^2$. Remarkably, the strongest deviations
from the data were found in the region $0.2 < |t| <
0.5$~(GeV/$c)^2$ which includes the transition between the
dominance of single and double scattering. The $pd$ elastic
observables (especially the tensor ones with mixed $x$ and $z$ polarization components) in this region are highly sensitive to the deuteron
quadrupole form factor. So, the origin of the observed discrepancies
should be sought in improper treatment of the deuteron $D$-wave component at short distances and in
probable contributions from the missing dynamical mechanisms associated with the short-range structure of the deuteron. A good candidate for such a mechanism is the proton scattering from the deuteron as a whole, i.e., from the short-range (dibaryon) component of the deuteron. In this case, one deals with the novel type of three-body force~\cite{Kukulin04-1,PKJPCS12}, the contribution of which to $pd$ scattering should rise with momentum transfer.


Since the validity of the Glauber model is restricted by the
momentum transfer, the angular range of its applicability becomes
larger at lower energies. This explains the very good agreement
between the refined Glauber model calculations and experiment for
the $pd$ elastic cross section and spin observables at $T_p =
135$~MeV till $\theta_{\rm c.m.} \simeq
80$~deg.~\cite{Temerbayev15}, while in the GeV region the data are
described well till $\theta_{\rm c.m.} \simeq 30$ deg. only.

The present study, along with earlier works, proves the refined
Glauber model to be a very useful tool for describing scattering of fast
protons off deuterons (and, in general, fast hadrons off
loosely-bound nuclei) at low momentum transfers $|t| \lesssim
0.2$--$0.3$~(GeV/$c)^2$ in a wide range of intermediate energies (at
least $T_p = 135$--$1135$~MeV). In this region, the exact
multiple-scattering series may be represented by the Glauber
amplitude based solely on the on-shell two-body interactions,
without need for solution of very complicated three-body
equations. However, the more accurate theoretical treatment is
needed at higher momentum transfers. It would also be highly
desirable to obtain more precise data for $pd$ elastic spin
observables with mixed $x$ and $z$ polarization components, which
presently have very large experimental uncertainties.

\acknowledgments The authors appreciate fruitful discussions with
Colin Wilkin, who carefully read the initial version of the manuscript and made valuable comments and suggestions. The work was supported by the Russian Foundation for
Basic Research, grants Nos.~19-02-00014 and 19-02-00011, and the Foundation for the Advancement of Theoretical Physics and Mathematics ``BASIS''.
\bibliography{pd-spin-obs}
\appendix*
\section{Transformation of polarization observables in $\bm{pd}$ elastic scattering to the Madison frame}

In this Appendix we give the explicit formulas which should be
applied to transform the $pd$ elastic spin observables from the
$xyz = \{\hat{q}\hat{n}\hat{k}\}$ frame used in deriving the
formalism of the refined Glauber model to the $xyz =
\{\hat{S}\hat{N}\hat{L}\}$ (Madison) frame conventionally used in
experiments (see definitions in Sec.~\ref{sec:model}).

There are the expressions for analyzing powers:
\begin{eqnarray}
\label{apm}
 A_y^{p\,({\rm Mad})} \!&=&\! -{A_y}^p, \quad A_y^{d\,({\rm Mad})} = -{A_y}^d, \nonumber \\
 A_{yy}^{({\rm Mad})} \!&=&\! A_{yy}, \nonumber \\
 A_{xx}^{({\rm Mad})} \!&=&\! \frac{1}{2}(1 + {\rm
cos}\theta)\,A_{xx} + \frac{1}{2}(1 - {\rm cos}\theta)\,A_{zz}
\nonumber \\
\!& &\! - {\rm sin}\theta\,A_{xz}, \nonumber \\
 A_{xz}^{({\rm Mad})} \!&=&\! -{\rm cos}\theta\,A_{xz} -
\frac{1}{2}{\rm sin}\theta\,(A_{xx} - A_{zz}), \nonumber \\
 A_{zz}^{({\rm Mad})} \!&=&\! -A_{yy}^{({\rm Mad})} - A_{xx}^{({\rm
 Mad})},
\end{eqnarray}
and for spin-correlation parameters:
\begin{eqnarray}
 \label{scpm}
 C_{y,y}^{({\rm Mad})} \!&=&\!  C_{y,y}, \nonumber \\
 C_{x,x}^{({\rm Mad})} \!&=&\! \frac{1}{2}(1 + {\rm cos}\theta)\,C_{x,x}
+ \frac{1}{2}(1-{\rm cos}\theta)\,C_{z,z} \nonumber \\
\!& &\! - \frac{1}{2}{\rm
sin}\theta\,(C_{z,x} + C_{x,z}), \nonumber \\
 C_{z,x}^{({\rm Mad})} \!&=&\! -\frac{1}{2}(1 + {\rm cos}\theta)\,C_{z,x}
+ \frac{1}{2}(1-{\rm cos}\theta)\,C_{x,z} \nonumber \\
\!& &\! - \frac{1}{2}{\rm
sin}\theta\,(C_{x,x} - C_{z,z}), \nonumber \\
 C_{x,z}^{({\rm Mad})} \!&=&\! -\frac{1}{2}(1 + {\rm cos}\theta)\,C_{x,z} + \frac{1}{2}(1-{\rm
 cos}\theta)\,C_{z,x} \nonumber \\
\!& &\! - \frac{1}{2}{\rm sin}\theta\,(C_{x,x} - C_{z,z}), \nonumber \\
 C_{z,z}^{({\rm Mad})} \!&=&\! \frac{1}{2}(1 + {\rm cos}\theta)\,C_{z,z} + \frac{1}{2}(1-{\rm
 cos}\theta)\,C_{x,x} \nonumber \\
\!& &\! + \frac{1}{2}{\rm sin}\theta\,(C_{x,z} + C_{z,x}), \nonumber \\
 C_{y,yy}^{({\rm Mad})} \!&=&\! -C_{y,yy}, \nonumber \\
 C_{y,xx}^{({\rm Mad})} \!&=&\! -\frac{1}{2}(1 + {\rm
cos}\theta)\,C_{y,xx} - \frac{1}{2}(1 - {\rm cos}\theta)\,C_{y,zz} \nonumber \\
\!& &\! + {\rm sin}\theta\,C_{y,xz}, \nonumber \\
 C_{y,xz}^{({\rm Mad})} \!&=&\! {\rm cos}\theta\,C_{y,xz} +
\frac{1}{2}{\rm sin}\theta\,(C_{y,xx} - C_{y,zz}), \nonumber \\
 C_{y,zz}^{({\rm Mad})} \!&=&\! -C_{y,yy}^{({\rm Mad})} - C_{y,xx}^{({\rm Mad})}, \nonumber \\
 C_{x,xy}^{({\rm Mad})} \!&=&\! -\frac{1}{2}(1 + {\rm cos}\theta)\,C_{x,xy}
- \frac{1}{2}(1-{\rm cos}\theta)\,C_{z,zy} \nonumber \\
\!& &\! + \frac{1}{2}{\rm sin}\theta\,(C_{z,xy} + C_{x,zy}), \nonumber \\
 C_{z,xy}^{({\rm Mad})} \!&=&\! \frac{1}{2}(1 + {\rm cos}\theta)\,C_{z,xy}
 - \frac{1}{2}(1-{\rm cos}\theta)\,C_{x,zy} \nonumber \\
\!& &\! + \frac{1}{2}{\rm sin}\theta\,(C_{x,xy} - C_{z,zy}), \nonumber \\
 C_{x,zy}^{({\rm Mad})} \!&=&\! \frac{1}{2}(1 + {\rm cos}\theta)\,C_{x,zy} - \frac{1}{2}(1-{\rm
 cos}\theta)\,C_{z,xy} \nonumber \\
\!& &\! + \frac{1}{2}{\rm sin}\theta\,(C_{x,xy} - C_{z,zy}), \nonumber \\
 C_{z,zy}^{({\rm Mad})} \!&=&\! -\frac{1}{2}(1 + {\rm cos}\theta)\,C_{z,zy} - \frac{1}{2}(1-{\rm
 cos}\theta)\,C_{x,xy} \nonumber \\
\!& &\! - \frac{1}{2}{\rm sin}\theta\,(C_{z,xy} + C_{x,zy}).
\end{eqnarray}

We should note here that we did not transform the analyzing powers
to the Madison frame in the previous works~\cite{PKPRC10,PKYAF10}.
However, when compared the theoretical predictions to experimental
data, we inverted the sign of vector analyzing powers $A_y^p$ and
$A_y^d$. We also considered tensor analyzing powers $A_{yy}$
(which is the same in both coordinate frames) and $A_{xx}$ (which
is changed only slightly by Eq.~(\ref{apm}) in the forward
hemisphere), so that, there was no significant error in comparing
these observables to those measured in the Madison frame. The only
analyzing power that changes its behavior substantially under the
transformation~(\ref{apm}) is $A_{xz}$ (due to an admixture of
large $A_{xx}$), which was not considered in our previous works.

The expressions for some of the above observables in terms of 18
$pd$ amplitudes derived directly in the Madison frame and the
relations of these amplitudes to our ones $A_1$--$A_{12}$ are to
be found in Ref.~\cite{Temerbayev15}.

We also give here an alternative representation of the $pd$
elastic scattering amplitude in the $\{\hat{q}\hat{n}\hat{k}\}$
frame, which has a more symmetric form than Eq.~(\ref{ma}):
\begin{eqnarray}
\label{mm}
   M[{\bf p},{\bf  q}; {\bm \sigma}, {\bf  S}] \!&=&\! \bigl(M_1 + M_2 \,{\bm
\sigma}\!\cdot\!\hat{n}\bigr)(1 - ({\bf S}\!\cdot\!\hat{q})^2) \nonumber \\
 & & + \bigl(M_3 + M_4
\,{\bm \sigma}\!\cdot\!\hat{n}\bigr)(1 - ({\bf S}\!\cdot\!\hat{n})^2) \nonumber \\
 & & + \bigl(M_5 + M_6 \,{\bm \sigma}\!\cdot\!\hat{n}\bigr)(1 - ({\bf S}\!\cdot\!\hat{k})^2) \nonumber \\
 & & + iM_7\,{\bm \sigma}\!\cdot\!\hat{k}\,{\bf S}\!\cdot\!\hat{k} \nonumber \\
 & & - M_{8}\,{\bm \sigma}\!\cdot\!\hat{q}({\bf S}\!\cdot\!\hat{q}\,{\bf
S}\!\cdot\!\hat{n} + {\bf S}\!\cdot\!\hat{n}\,{\bf S}\!\cdot\!\hat{q}) \nonumber \\
 & & - i\bigl(M_9 + M_{10}\,{\bm
\sigma}\!\cdot\!\hat{n}\bigr){\bf S}\!\cdot\!\hat{n} +
iM_{11} \,{\bm \sigma}\!\cdot\!\hat{q}\,{\bf S}\!\cdot\!\hat{q} \nonumber \\
 & & - M_{12} \,{\bm \sigma}\!\cdot\!\hat{k}({\bf S}
\!\cdot\!\hat{k}\,{\bf S}\!\cdot\!\hat{n} + {\bf
S}\!\cdot\!\hat{n}\,{\bf S}\!\cdot\!\hat{k}).
\end{eqnarray}


The set of invariant amplitudes $M_1$--$M_{12}$ (multiplied by a standard normalization factor $8p\sqrt{\pi s}$) was used
in, e.g., Ref.~\cite{Uzikov98} (note however that the direction of
the unit vector $\hat{n}$ was chosen there opposite to ours). These
amplitudes are related to our ones $A_1$--$A_{12}$ as follows:
\begin{eqnarray}
A_1 &=& M_1 + M_3 - M_5, \quad A_2 = M_2 + M_4 - M_6, \nonumber \\
A_3 &=& M_5 - M_1, \quad A_4 = M_6 - M_2, \nonumber \\
A_5 &=& M_5 - M_3, \quad A_6 = M_6 - M_4, \nonumber \\
A_7 &=& iM_7, \quad A_8 = -M_8, \quad  A_9 = -iM_9, \nonumber \\
A_{10} &=& -iM_{10}, \quad A_{11} = iM_{11}, \quad A_{12} = -M_{12}.
\end{eqnarray}
The $pd$ elastic observables expressed in terms of the amplitudes
$M_1$--$M_{12}$ look simpler than those given by Eq.~(4) of
Ref.~\cite{PKPRC10} and Eq.~(\ref{scp}) of the present paper. In
particular, interference between different amplitudes vanishes in
the expression for the differential cross section.
On the other hand, such a representation is less transparent in
sense that there are three large amplitudes $M_1$, $M_3$ and $M_5$
instead of only one dominant amplitude $A_1$ at low momentum
transfers.

\end{document}